\documentclass[a4paper,twocolumn,11pt]{quantumarticle}
\pdfoutput=1
\usepackage[utf8]{inputenc}
\usepackage[english]{babel}
\usepackage[T1]{fontenc}
\usepackage{amsmath}
\usepackage[colorlinks,breaklinks,linkcolor=blue,anchorcolor=blue,citecolor=blue,urlcolor=blue]{hyperref}
\usepackage{subfigure}
\usepackage{tikz}
\usepackage{lipsum}
\usepackage{epstopdf}
\usepackage{soul}
\begin{document}

\title{Controlled unidirectional reflectionlessness by coupling strength in a non-Hermitian waveguide quantum electrodynamics system}
\author{QiuDe-xiu }
\affiliation{Center for Quantum Sciences and School of Physics, Northeast Normal University, Changchun 130024, China}
\author{LiFude }
\orcid{0000-0003-0290-4698}
\affiliation{Center for Quantum Sciences and School of Physics, Northeast Normal University, Changchun 130024, China}
\author{XueKang }
\affiliation{Center for Quantum Sciences and School of Physics, Northeast Normal University, Changchun 130024, China}
\author{YiXuexi}
\affiliation{Center for Quantum Sciences and School of Physics, Northeast Normal University, Changchun 130024, China}
\email{yixx@nenu.edu.cn}
\maketitle

\begin{abstract}
  Unidirectional reflectionlessness is investigated in a waveguide quantum electrodynamics system that consists of a cavity and a $\Lambda$-type three-level quantum dot coupled to a one-dimensional plasmonic waveguide. Analytical expressions of transmission and reflection coefficients are derived and discussed for both resonant and off-resonant couplings. By appropriately modulating the coupling strength, phase shift and Rabi frequency, unidirectional reflectionlessness is observed  at the exceptional points. And unidirectional coherent perfect absorption is exhibited at the vicinity of exceptional point. These results might find applications in designing quantum devices of photons, such as optical switches and single-photon transistors.
\end{abstract}

The discovery of Carl Bender \emph{et al.} in 1998\cite{BenderCM1998} was a crucial milestone in non-Hermitian systems,
that non-Hermitian Hamiltonian with parity-time symmetry can show entire real eigenvalue spectra.
This has attracted considerable attention to explore the systems with such feature at exceptional points (EPs)\cite{JingH2017,PengB2014Science,PengB2016,DopplerJ2016}.
The various exotic phenomena are developed at the EPs,
such as nonreciprocal light propagation\cite{PengB2014,ChangL2014,HeB2018,XiaK2018},
optical isolation\cite{RegensburgerA2012},
optomechanically induced transparency\cite{JingH2015},
coherent absorption\cite{BaumB2015,HuangCY2014,SunY2014},
and unidirectional reflectionlessness\cite{LinZ2011,AlaeianH2014,GearJ2015,FuY2016,FengL2013,YangE2016,ShenY2014,
Gu2017OE,Bai2017SR,Bai2017APE,Yin2018OC,Liu2018JNOPM,HuangY2016,ZhangC2017,ZhangC2017IEEE,ZhangC2017OE}, and so on.

With the strict requirements for the minimization of the operating platform,
the architecture of one-dimensional (1D) waveguides coupled to quantum emitters with inherent advantages
in quantum electrodynamics (QED) system has attracted widespread attention in recent years.
Quantum emitters include
single or multiple atoms\cite{ShenJT2007,Fan Shanhui2010,Rephaeli Eden2011,Longo Paolo2011,Zheng Huaixiu2012,Roy Dibyendu2013},
quantum dots (QDs)\cite{Englund Dirk2010,MTChen2012OL,Huang Jin-Feng2013},
optomechanical cavities\cite{Liao Jie-Qiao2013},
cavity-atom\cite{Shi T2011},
whispering-gallery resonator\cite{JTshen2009PRA} and the others.
These 1D waveguide systems are excellent platforms for studying the photon-matter interaction
and controlling the photon scattering properties.
It has brought out a significant advance, e.g.,
switches\cite{Yan C2011,Shomroni2014},
transistors\cite{Witthaut2010},
routers\cite{Gonzalez-Ballestero Carlos2016,Yan Cong-Hua2018},
electromagnetic induced transparency (EIT)\cite{RoyD2011},
non-reciprocal photon scattering\cite{Xu2017,Shangcheng2019,Nie2021}.

Since unidirectional reflectionlessness as part of non-reciprocal transport has been experimentally demonstrated\cite{Huang2015},
it has been expanded from optical waveguide systems\cite{Gu2017OE,Bai2017SR,Bai2017APE} to surface plasmonic waveguide systems\cite{Wu2018OE,Qiu2019QIP1,Qiu2019QIP2,Yang2020QST}.
Based on such non-Hermitian waveguide systems,
Wu \emph{et al.} demonstrated the single-band and dual-band unidirectional reflectionlessness of surface plasmon
in a system of the multiple QDs coupled to a plasmonic waveguide\cite{Wu2018OE}.
Qiu \emph{et al.} investigated dual-band unidirectional reflectionlessness and non-reciprocal entanglement
in a non-Hermitian plasmonic cavities-waveguide system\cite{Qiu2019QIP1,Qiu2019QIP2}.
Yang \emph{et al.} studied single-band and dual-band unidirectional reflectionless phenomena at EPs
by advisably regulating classical driving field in a non-Hermitian quantum system with two $\Lambda$-type three-level QDs\cite{Yang2020QST}.
Moreover, inspired by these proposals,
it is worth noticing to achieve unidirectional refltectionlessness
by controlling coupling strength between two short-distance emitters for two directions in a waveguide QED system.

In this paper, we propose a non-Hermitian waveguide QED system to realize unidirectional refltectionlessness
by utilizing the coupling strength between a cavity and a $\Lambda$-type three-level QD.
When the $\Lambda$-type three-level QD is driven by classical driving field,
we show the dual-band unidirectional reflectionlessness in reflection spectra.
In the absence of classical driving field,
we demonstrate single-band unidirectional reflectionlessness and unidirectional coherent perfect absorption
in the case of the three-level QD being degenerated to two-level QD.

\section{Model and calculation}\label{Sec:Model and calculation}
\begin{figure}[h]
\centering
\includegraphics[width=8cm]{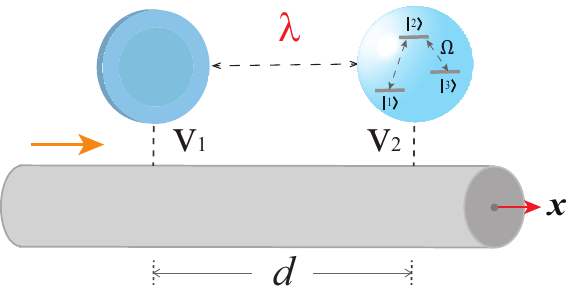}
\caption{Schematic diagram of a plasmonic waveguide coupled to a cavity and a $\Lambda$-type three-level QD with a distance $d$ and coupling strength $\lambda$.} \label{01}
\end{figure}
Figure 1 shows a waveguide quantum electrodynamics system,
where a cavity and a $\Lambda$-type three-level QD side coupled to
a plasmonic waveguide with coupling strength $\lambda$ and the distance $d$.
The transition of QD $|1\rangle\leftrightarrow|2\rangle$ is coupled to the plasmonic waveguide with coupling strength $V_{2}$,
nevertheless, the transition of QD $|2\rangle\leftrightarrow|3\rangle$ is driven by a classical driving field with Rabi frequency $\Omega$.

The effective non-Hermitian Hamiltonian (H) of system in the real space under the rotating wave approximation (assuming $\hbar=1$) can be derived as\cite{JTshen2009PRA,PRB842011}
\begin{small}
\begin{eqnarray}\label{e1}
H&=&\int dx\{-iv_{g}c^{\dag}_{R}(x)\frac{d}{dx}c_{R}(x)+iv_{g}c^{\dag}_{L}(x)\frac{d}{dx}c_{L}(x)\}\nonumber\\
&&+(\omega_{1}-i\gamma_{1})a^{\dag}a+(\omega_{2}-i\gamma_{2})\sigma_{22}\nonumber\\
&&+(\omega_{2}-\Delta_{23}-i\gamma_{3})\sigma_{33}+\omega_{0}\sigma_{11}+\frac{\Omega}{2}(\sigma_{23}+\sigma_{32})\nonumber\\
&&+V_{1}\int dx\delta(x)[c^{\dag}_{R}(x)a+c^{\dag}_{L}(x)a+ H.c.]\nonumber\\
&&+V_{2}\int dx\delta(x-d)[c^{\dag}_{R}(x)\sigma_{12}+c^{\dag}_{L}(x)\sigma_{12}+ H.c.]\nonumber\\
&&+\lambda(a^{\dag}\sigma_{12}+a\sigma_{21}),
\end{eqnarray}
\end{small}
where
$c^{\dag}(x)$ ($c(x)$) is the Fourier transform of the bosonic creation (annihilation) operator,
describing the left- and right-traveling surface plasmon along waveguide at position $x$.
$v_{g}$ is the group velocity of incident photon.
$\omega_{1}$ and $\gamma_{1}$ are resonance frequency and dissipation of the cavity, respectively.
We take the state $|1\rangle$ as the origin of energy, $\omega_{0}=0$.
So, $\omega_{2}$ represents the transition frequency between $|1\rangle$ and $|2\rangle$.
$\gamma_{2}$ and $\gamma_{3}$ are dissipation of state $|2\rangle$ and $|3\rangle$, respectively.
$\Delta_{23}$ denote detuning between energies of classical driving field and transition of the QD $|2\rangle\leftrightarrow|3\rangle$.
$V_{1(2)}$ is coupling strength between the plasmonic waveguide and the cavity (QD).
For simplicity, one suppose that the coupling strength $V=V_{1}=V_{2}$ in the full.
$a^{\dag}$ and $a$ are creation and annihilation operators of the cavity
and $\sigma_{mn}$ represents the dipole transition operator between the states $|m\rangle$ and $|n\rangle$ ($m$, $n$=1,2,3).
$\lambda$ is the coupling strength between cavity and the $\Lambda$-type three-level QD.

Hence, the eigenstate of system can be expressed as
\begin{small}
\begin{eqnarray}\label{e2}
|E_{k}\rangle&=&\int dx[\psi_{R}(x)c^{\dag}_{R}(x)+\psi_{L}(x)c^{\dag}_{L}(x)]|{\o}\rangle|1\rangle\nonumber \\
&&+\xi_{1}a^{\dag}|{\o}\rangle|1\rangle+\xi_{2}|{\o}\rangle|2\rangle+\xi_{3}|{\o}\rangle|3\rangle,
\end{eqnarray}
\end{small}
where $\xi_{j}$ ($j$=1,2,3) is the excitation amplitude of the cavity field and the $\Lambda$-type three-level QD.
$|{\o}\rangle|1\rangle$ denotes that cavity is in the vacuum and the QD is in ground state $|1\rangle$ without photon in the waveguide.
Assuming that the incident surface plasmon is coming from the left which defined as the forward direction with the energy $E_{k}=\hbar\omega_{k}$.
Here, the scattering amplitudes $\psi_{R}(x)$ and $\psi_{L}(x)$ can be expressed as
\begin{small}
\begin{eqnarray}\label{e3}
\psi_{R}(x)&=& e^{ikx}[F(-x)+a F(x)F(d-x)+t F(x-d)],\nonumber\\
\psi_{L}(x)&=& e^{-ikx}[r F(-x)+b F(x) F(d-x)],
\end{eqnarray}
\end{small}
where $F(x)$ is the unit step function, $t$ and $r$ are the transmission and reflection amplitudes, respectively.
$e^{ikx}aF(x)F(x-d)$ and $e^{-ikx}bF(x) F(d-x)$ represent the wave function of the surface plasmon between the cavity and QD.

Then, by solving the eigenvalue equation $H|E_{k}\rangle = E_{k}|E_{k}\rangle$, one can derive the following relations
\begin{small}
\begin{eqnarray}\label{e4}
(\Delta_{2} + i\gamma_{2})\xi_{2}&=&\lambda \xi_{1} +\frac{\Omega}{2}\xi_{3}+ V_{2}[(a + t_{f}) e^{i\theta} + b e^{-i\theta}],\nonumber \\
 (\Delta_{1} + i\gamma_{1})\xi_{1}&=&\lambda \xi_{2} + V_{1}(1 + a + r_{f} + b), \nonumber\\
 \frac{\Omega}{2}\xi_{2} &=& (\Delta_{3} + i\gamma_{3})\xi_{3},\nonumber\\
 a &=& 1 + \frac{V_{1} \xi_{1}}{iv_{g}},\quad t_{f} = a + \frac{V_{2}\xi_{2}}{iv_{g}}e^{-i \theta},\nonumber\\
 r_{f} &=& b + \frac{V_{1}\xi_{1}}{iv_{g}}, \quad b = \frac{V_{2} \xi_{2}}{iv_{g}}e^{i \theta},
\end{eqnarray}
\end{small}
here, $\Delta_{j}$ is the detuning between the incident surface plasmon with the energy $E_{k}$ and the cavity resonance frequency $\omega_{1}$ and the $\Lambda$-type three-level QD with energy-level frequency $\omega_{2}$ and $\omega_{3}$, $\Delta_{j}= \omega-\omega_{j}$.
$\theta=kd$ is the phase shift between the cavity and QD.

By solving the above system of equation, we can obtain the transmission amplitude $t_{f}$ and the reflection amplitude $r_{f}$ for forward direction.
And the corresponding transmission amplitude $t_{b}$ and reflection amplitude $r_{b}$ when the surface plasmon is coming in from the right side
which defined as the backward direction can also be given.
So, the transmission amplitude and reflection amplitude for forward and backward directions as follows
\begin{widetext}
\begin{eqnarray}\label{e5}
t&=&t_{f}=t_{b}\frac{  e^{-i\theta}\{ 4\Gamma\lambda(-1+e^{2i\theta})B_{3}- e^{i\theta}[ 4\gamma_{3}(\lambda^{2}-\Delta_{1}B_{2})+ A -i( 4\lambda^{2}\Delta_{3}+\Delta_{1}( \Omega^{2}-4\Delta_{3}B_{2}))]\}
}{P},\\
r_{f}&=&-\frac{C-\Gamma\{\Omega^{2}+4\gamma_{3}[e^{2i\theta}\gamma_{1}+\gamma_{2}-i(2e^{i\theta}\lambda+e^{2i\theta}\Delta_{1}+\Delta_{2})
+4\Delta_{3}(2e^{i\theta}\lambda+e^{2i\theta}B_{1}+B_{2})]\}
}{P},\\
r_{b}&=&-\frac{C-\Gamma\{e^{2i\theta}\Omega^{2}+4\gamma_{3}[\gamma_{1}+e^{2i\theta}\gamma_{2}-i(2e^{i\theta}\lambda+\Delta_{1}+e^{2i\theta}\Delta_{2})
+4\Delta_{3}(2e^{i\theta}\lambda+B_{1}+e^{2i\theta}B_{2})]\}
}{P},
\end{eqnarray}
\end{widetext}
where,
\begin{small}
\begin{eqnarray}
A &\equiv& \gamma_{1}[\Omega^{2}-4i\gamma_{3}\Delta_{2}+4\gamma_{2}(\gamma_{3}-i\Delta_{3})-4\Delta_{2}\Delta_{3}],\nonumber\\
B_{1} &\equiv& i\gamma_{1}+\Delta_{1},\quad B_{2} \equiv i\gamma_{2}+\Delta_{2},\quad B_{3} \equiv i\gamma_{3}+\Delta_{3},\nonumber\\
C &\equiv& 4\Gamma^{2}(-1+e^{2i\theta})(\gamma_{3}-i\Delta_{3}),\nonumber
\end{eqnarray}
\end{small}
\begin{small}
\begin{eqnarray}
P&\equiv& C-\Gamma\{\Omega^{2}+4\gamma_{3}[\gamma_{1}+\gamma_{2}-i(2e^{i\theta}\lambda+\Delta_{1}+\Delta_{2})]\nonumber\\
&&-4\Delta_{3}(2e^{i\theta}\lambda+i\gamma_{1}+i\gamma_{2}+\Delta_{1}+\Delta_{2})\}\nonumber\\
&&+\{i\Omega^{2}\Delta_{1}+\gamma_{3}(-4\lambda^{2}+4i\gamma_{2}\Delta_{1}+4\Delta_{1}\Delta_{2})\nonumber\\
&&+4i\lambda^{2}\Delta_{3}+4\gamma_{2}\Delta_{1}\Delta_{3}-4i\Delta_{1}\Delta_{2}\Delta_{3}-A\}.\nonumber
\end{eqnarray}
\end{small}
$\Gamma=2V^{2}/v_{g}$ is the decay rate of the cavity or QD into the waveguide.
$t_{f}=t_{b}$
Therefore, the transmission $T=|t|^{2}$, the reflection $R_{f}$=$|r_{f}|^{2}$ and $R_{b}$=$|r_{b}|^{2}$  can be obtained to explore novel reflection phenomena.
Furthermore, the  absorption for forward ($A_{f}$) and backward ($A_{b}$) directions can be expressed as $A_{f} = 1 - T - R_{f}$ and $A_{b} = 1 - T - R_{b}$.

\section{Results and Discussion}\label{Sec:Results and Discussion}

In this section,
we first discuss the situation of a cavity and a $\Lambda$-type three-level QD with driving field coupled to a plasmonic waveguide.
According to Eqs. (5)-(7),
the reflection as a function of coupling strength $\lambda$ and detuning $\Delta_{1}$ for forward and backward directions is plotted in Fig. 2.
\begin{figure}[h]
\centering
\includegraphics[angle=0,width=0.5\textwidth]{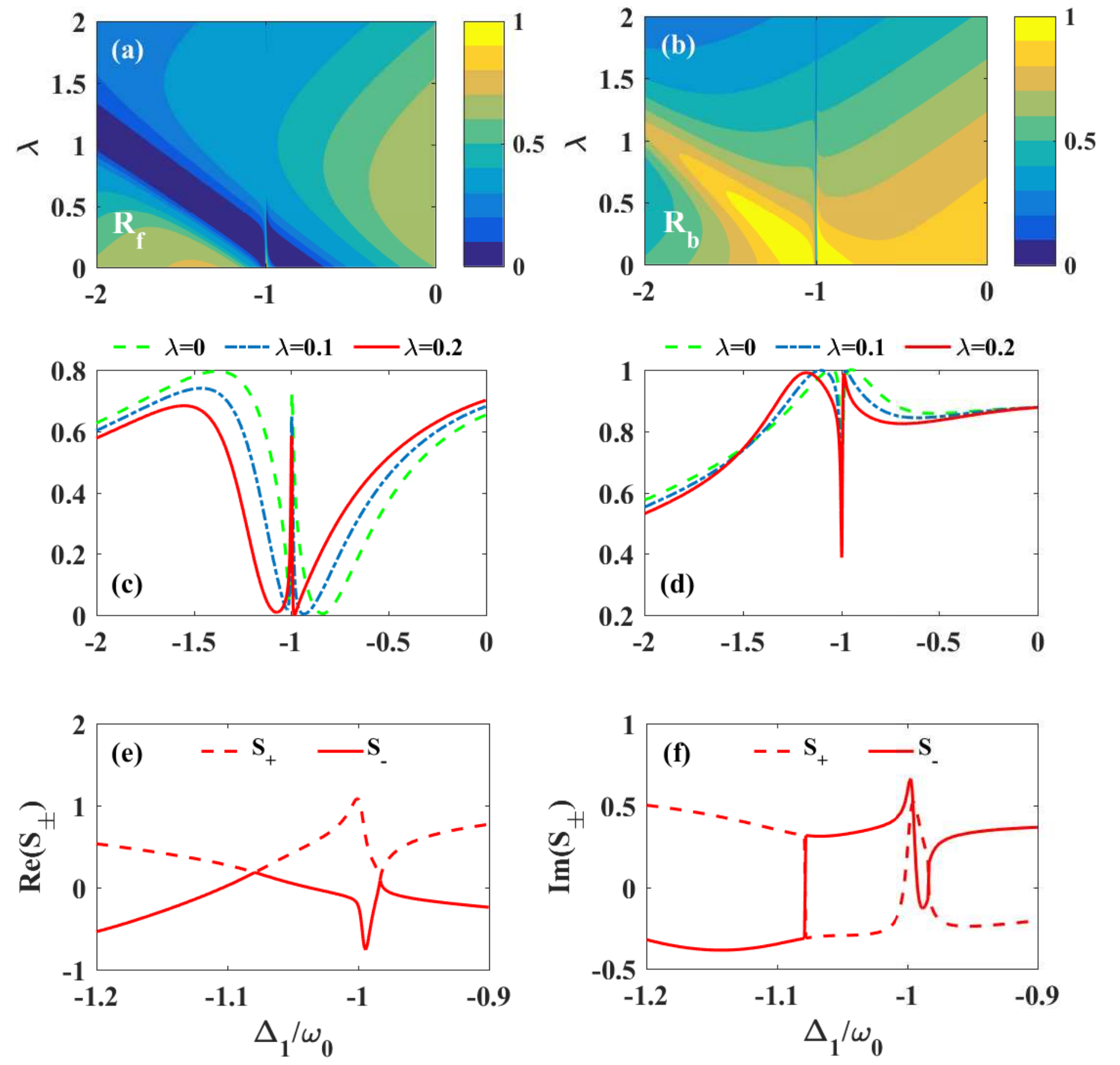}
\caption{(Color online)  (a)-(d) Reflection for forward direction $R_{f}$ and backward direction $R_{b}$ as a function of coupling strength $\lambda$ and detuning $\Delta_{1}$. (e) and (f) real and imaginary parts of eigenvalues $s_{\pm}$ versus $\Delta_{1}$. Red dashed lines and solid lines represent the eigenvalues $s_{+}$ and $s_{-}$, respectively. (e)-(f) $\lambda=0.2$ and other parameters are $\theta=0.1\pi$, $\Omega=0.1$, $\Gamma= 1.2$, $\gamma_{1} = 0.27$ and $\gamma_{2} = 0.001$ in units of $\omega_{0}=1\times10^{14}$.} \label{002}
\end{figure}
From Figs. 2(a) and 2(b),
it is shown that the reflection spectra for forward and backward directions
have red-shifts as $\lambda$ increases.
The red shift is clearly shown for several special values in Figs. 2(c) and 2(d).
There is a distinct fault at $\Delta_{1}=-1$ due to the transition $|2\rangle\leftrightarrow|3\rangle$, which causes the reflection spectra to have two extreme values.
That is,
the two low reflection regions (Fig. 2(a)) for forward direction correspond to the high reflection regions (Fig. 2(b)) for backward direction
in coupling strength range of 0 $\sim$ 0.61 and 0 $\sim$ 0.37.
Therefore,
the dual-band unidirectional reflectionlessness phenomena can be observed in the non-Hermitian quantum system.

To illuminate this unique phenomenon,
we will discuss further by scattering matrix $S$ of the system.
Based on Eqs. (5)-(7), it can be written as\cite{104}
\begin{eqnarray}
    S=
\left(%
\begin{array}{cc}
  t & r_{b} \\
  r_{f} & t \\
\end{array}
\right),
\end{eqnarray}
the eigenvalues of the scattering matrix $S$ can be expressed as
\begin{eqnarray}\label{e5}
    s_{\pm} = t \pm \sqrt{r_{f}r_{b}}.
\end{eqnarray}
There, if $\sqrt{r_{f}r_{b}}$ is equal to zero ($r_{f}=0$, $r_{b}\neq0$ or $r_{f}\neq0$, $r_{b}=0$), two eigenvalues coalesce and EP appears.
Namely, unidirectional reflectionlessness appears at the EP.
In Figs. 2(e) and 2(f),
we can see that
the real and imaginary parts of two eigenvalues $s_{+}$ and $s_{-}$ coalesce and cross
at detuning $\Delta_{1}=-1.08$ and $\Delta_{1}=-0.98$ when phase shift $\theta=0.1\pi$, respectively.
In other words,
we can obtain the same complex eigenvalues $s_{+}=s_{-}$ and two EPs occur.
In this case, $t$ is complex and $\sqrt{r_{f}r_{b}}=0$.
Hence, the dual-band unidirectional reflectionlessness phenomena can be obtained at two EPs.

As shown in Fig. 3,
one study the influences of Rabi frequency $\Omega$ and phase shift $\theta$ on reflections and absorptions for forward and backward directions, respectively.
From Figs. 3(a) and 3(b),
the near-zero reflection regions and high reflection regions for forward and backward directions are gradually separated
as $\Omega$ increases when phase shift $\theta=0.1\pi$.
This moment,
the low reflection regions for forward direction (Fig. 3(a)) correspond to the high reflection regions for forward direction (Fig. 3(b))
in the range of Rabi frequency from 0 to 0.9.
Meanwhile,
the high absorptivity for forward direction can be got in wide range of Rabi frequency 0$\sim$0.38 and 0$\sim$0.73.
Then, the dependencies of phase shift on reflection for forward and backward directions as shown in Figs. 3(e) and 3(d).
It is obvious that
the low reflection regions (blue regions) for forward direction corresponding to the high reflection regions (yellow regions) for backward direction in the phase shift range of 0$\sim $0.22$\pi$ and 1.09$\pi$$\sim$1.38$\pi$.
However,the low reflection regions (Fig. 3(d)) for backward direction corresponding to the high reflection regions (Fig. 3(e)) for forward direction in the phase shift range
of 0.85$\pi$$\sim $0.96$\pi$ and 1.83$\pi$$\sim$1.97$\pi$.
At the same time,
the periodic high absorption can also be obtained in Figs. 3(g) and 3(h).
Therefore,
the dual-band unidirectional reflectionlessness and perfect absorption can be achieved by adjusting
Rabi frequency and phase shift.
\begin{figure}[h]
\centering
\includegraphics[angle=0,width=0.5\textwidth]{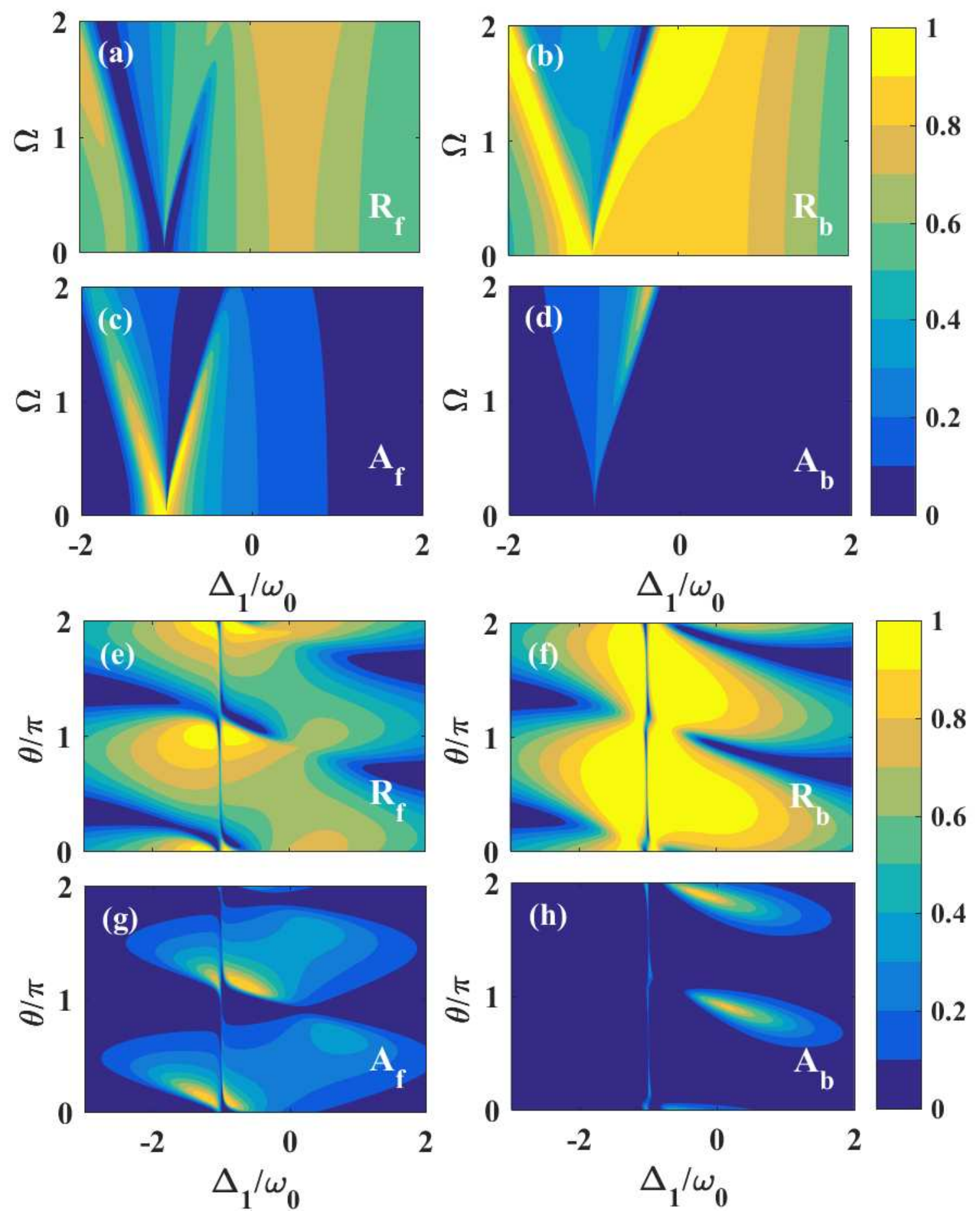}
\caption{(Color online) (a)-(d) Reflections and absorptions as a function of Rabi frequency $\Omega$ and detuning $\Delta_{1}$ for forward and backward directions. (e)-(h) Reflections and absorptions as a function of phase shift $\theta$ and detuning $\Delta_{1}$ for forward and backward directions. (a)-(d) $\theta=0.1\pi$ (e)-(h) $\Omega=0.3$ and other parameters are same as in Fig. 2.} \label{03}
\end{figure}

Whereafter,
we consider the results for whether or not incident surface plasmon resonate with cavity and QD to explore unidirectional-reflectionlessness behaviours when $\Omega=0$.
It is noted that the $\Lambda$-type three-level QD back to the two-level QD when switching off the driving field.
Firstly,
we discuss the case that cavity and QD have same transition frequency $\omega_{1}=\omega_{2}$.
For convenience, the equal detunings in this section are uniformly marked as $\Delta$.
The transmission and reflection coefficients are now sorted as
\begin{widetext}
\begin{eqnarray}\label{e7}
t&=&\frac{e^{-i\theta}\{i(-1+e^{2i\theta})\lambda\Gamma+ e^{i\theta} [(\Delta + i \gamma_{1})(\Delta + i \gamma_{2})-\lambda^{2}]\}}{(-1+e^{2i\theta})\Gamma^{2}+i[2e^{i\theta}\lambda\Gamma+\Gamma(\Delta + i \gamma_{1})+\Gamma(\Delta + i \gamma_{2})]+[(\Delta + i \gamma_{1})(\Delta + i \gamma_{2})-\lambda^{2}]},\\
r_{f}&=&-\frac{(-1+e^{2i\theta})\Gamma^{2}+i[2e^{i\theta}\lambda\Gamma+e^{2i\theta}\Gamma(\Delta + i \gamma_{1})+\Gamma(\Delta + i \gamma_{2})]}
{(-1+e^{2i\theta})\Gamma^{2}+i[2e^{i\theta}\lambda\Gamma+\Gamma(\Delta + i \gamma_{1})+\Gamma(\Delta + i \gamma_{2})]+[(\Delta + i \gamma_{1})(\Delta + i \gamma_{2})-\lambda^{2}]},\\
r_{b}&=&-\frac{(-1+e^{2i\theta})\Gamma^{2}+i[2e^{i\theta}\lambda\Gamma+\Gamma(\Delta + i \gamma_{1})+e^{2i\theta}\Gamma(\Delta + i \gamma_{2})]}
{(-1+e^{2i\theta})\Gamma^{2}+i[2e^{i\theta}\lambda\Gamma+\Gamma(\Delta + i \gamma_{1})+\Gamma(\Delta + i \gamma_{2})]+[(\Delta + i \gamma_{1})(\Delta + i \gamma_{2})-\lambda^{2}]},
\end{eqnarray}
\end{widetext}
it exhibits clearly that
the coupling strength $\lambda$ and phase shift $\theta$ play the critical role to control reflection spectra $r_{f}$ and $r_{b}$.
\begin{figure}[h]
\centering
\includegraphics[angle=0,width=0.5\textwidth]{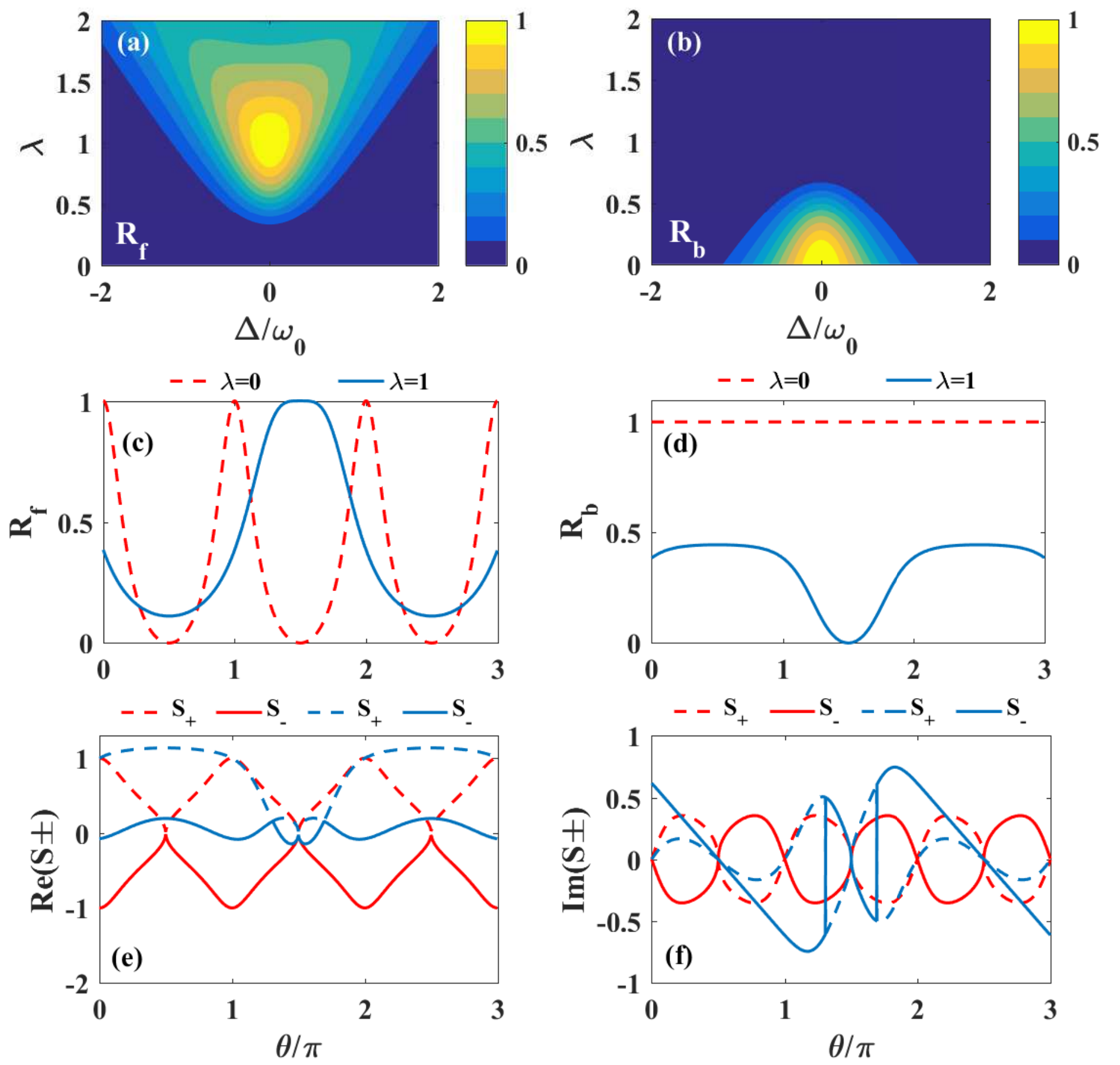}
\caption{(Color online)  (a) and (b) Reflection for forward direction $R_{f}$ and backward direction $R_{f}$ as a functions of coupling strength $\lambda$ and detuning $\Delta$, respectively, when the phase shift $\theta=1.5\pi$. (c) and (d) reflection spectra for forward and backward directions as a function of the phase shift $\theta$ for different coupling strength $\lambda=0$ and $1$, respectively. And (e) and (f) real and imaginary parts of eigenvalues $s_{\pm}$ as a function of the phase shift $\theta$ for different coupling strength $0$ (red lines) and $1$ (blue lines), respectively.
Other parameters are $\Gamma= 0.5$, $\gamma_{1} =1 $, $\gamma_{2} = 0.01$.} \label{002}
\end{figure}

Figure 4 shows
the reflection spectra versus detuning $\Delta$ and coupling strength $\lambda$
and the real and imaginary part curves of eigenvalues $s_{\pm}$ versus the phase shift $\theta$
for different $\lambda$.
From Figs. 4(a) and 4(b),
we observe that the high reflection region (yellow region) for forward direction (Fig. 4(a))
corresponds to the low reflection region (blue region) for backward direction (Fig. 4(b)).
Whereas, high reflection region for backward direction (Fig. 4(b))
corresponds to the low reflection region for forward direction (Fig. 4(a)).
That is to say,
the unidirectional reflectionlessness phenomena occur
in coupling strength regions of $0 \sim 0.2$ and $0.79 \sim 1.26$ around the detuning $\Delta=0$, respectively.
To better understand above phenomenon,
one plot the reflection spectra as a function of phase shift for different coupling strength
when the incident surface plasmon resonates with the cavity and QD in Figs. 4(c) and 4(d).
Equations (11) and (12) exhibit clearly that
the coupling strength $\lambda$ and phase shift $\theta$ play the critical role to control reflection spectra $r_{f}$ and $r_{b}$.
If $\lambda=0$, $R_{f}\approx0$ and $R_{b}\approx1$ at $\theta=(2n+1)\pi/2$ (n is an integer).
However, if $\lambda\neq0$, $R_{f}=1$ and $R_{b}=0$ at $\theta=(2m+1)\pi/2$ (m is an odd number).
The incident resonant surface plasmon for forward direction is almost completely reflected and the other direction is completely absorbed.
The real and imaginary part curves of eigenvalues $s_{\pm}$
can better explain the the deep-level physical mechanism of the unidirectional reflectionlessness in Figs. 4(e) and 4(f).
The real and imaginary parts of two eigenvalues $s_{+}$ and $s_{-}$ coalesce and cross
at phase shift $\theta = 0.15\pi$ when the coupling strength $\lambda=0$ and $\lambda=1$.
At this point, the same complex eigenvalue $s_{+}=s_{-} $ and there is only one real part and one imaginary part.
Notably, EP occurs at phase shift $\theta = 0.15\pi$, which $t$ is complex and $\sqrt{r_{f}r_{b}}$ is 0.
Therefore, unidirectional reflectionless is realized at EP by appropriately adjusting the coupling strength.
\begin{figure}[h]
\centering
\includegraphics[angle=0,width=0.5\textwidth]{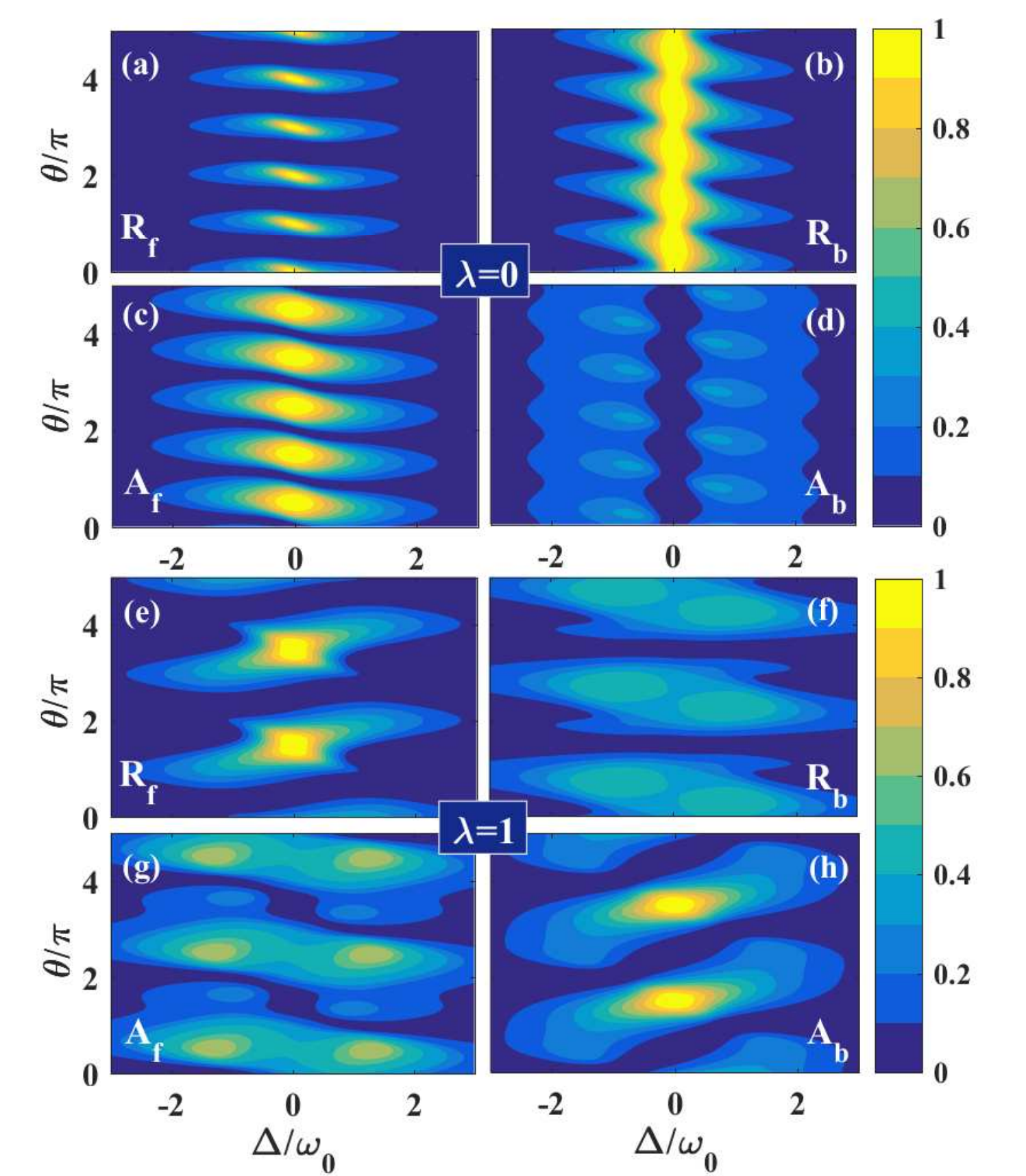}
\caption{(Color online)  Dependence of reflections and absorptions for forward and backward directions on phase shift $\theta$ and detuning $\Delta$ for different coupling strength $\lambda=0$ and $1$, respectively. These parameters are same as Fig. 2.} \label{003}
\end{figure}

Figure 5 exhibits the influences of phase shift $\theta$ and detuning $\Delta$ on reflections and absorptions for forward and backward directions, respectively.
From Figs. 5(a) and 5(b) and 5(e) and 5(f),
one can easily see that reflections for forward and backward directions vary periodically as increasing the phase shift $\theta$.
And the unidirectional reflectionlessness appears in the phase shift range from 0.31 to 0.69 and from 1.31 to 1.69, and from 1.27 to 1.7 and from 3.27 to 3.7.
Meanwhile,
the perfect absorption occurs in the case of forward incidence when coupling strength $\lambda=0$
and the largest absorption can be observed in the phase shift range from 0.32 to 0.68 and form 1.32 to 1.68 at detuning $\Delta=0$ as depicted in Fig. 5(c).
However, when coupling strength $\lambda=1$, the perfect absorption exists in the case of backward incidence
and the largest absorption is mainly concentrated in the range of phase shift 1.33$\sim$1.67 and 3.33$\sim$3.67.
Therefore, unidirectional reflectionlessness at EPs can be implemented over a wide phase shift range
and perfect absorption can also be flexibly regulated by varying the coupling strength.

Furthermore, for the phase shift $\theta_{1(2)}$ of cavity (QD),
the phase differences between cavity and QD for forward and backward directions are $\theta_{f}$=$\theta_{1}$ $-$ $\theta_{2}$ + $2\theta$
and $\theta_{b}$=$\theta_{2}$ $-$ $\theta_{1}$ + $2\theta$, respectively,
$\theta_{1(2)} = (\omega_{k} - \omega_{1(2)})/(\eta + \gamma_{1(2)}/2)$\cite{JTShen2005OL,HZheng2013PRL,FFratini2014PRL}.
The phase differences for forward (backward) and backward (forward) directions
at $\theta=0.1\pi$ ($0.9\pi$) are $\sim2\pi$ ($\sim\pi$) and $\sim\pi$ ($\sim2\pi$),
result in reflections for forward and backward directions at detuning $\Delta =-0.1$ and $\Delta =1.2$ are nearly 0 based on Fabry-P$\acute{e}$rot resonance.
\begin{figure}[h]
\centering
\includegraphics[angle=0,width=0.5\textwidth]{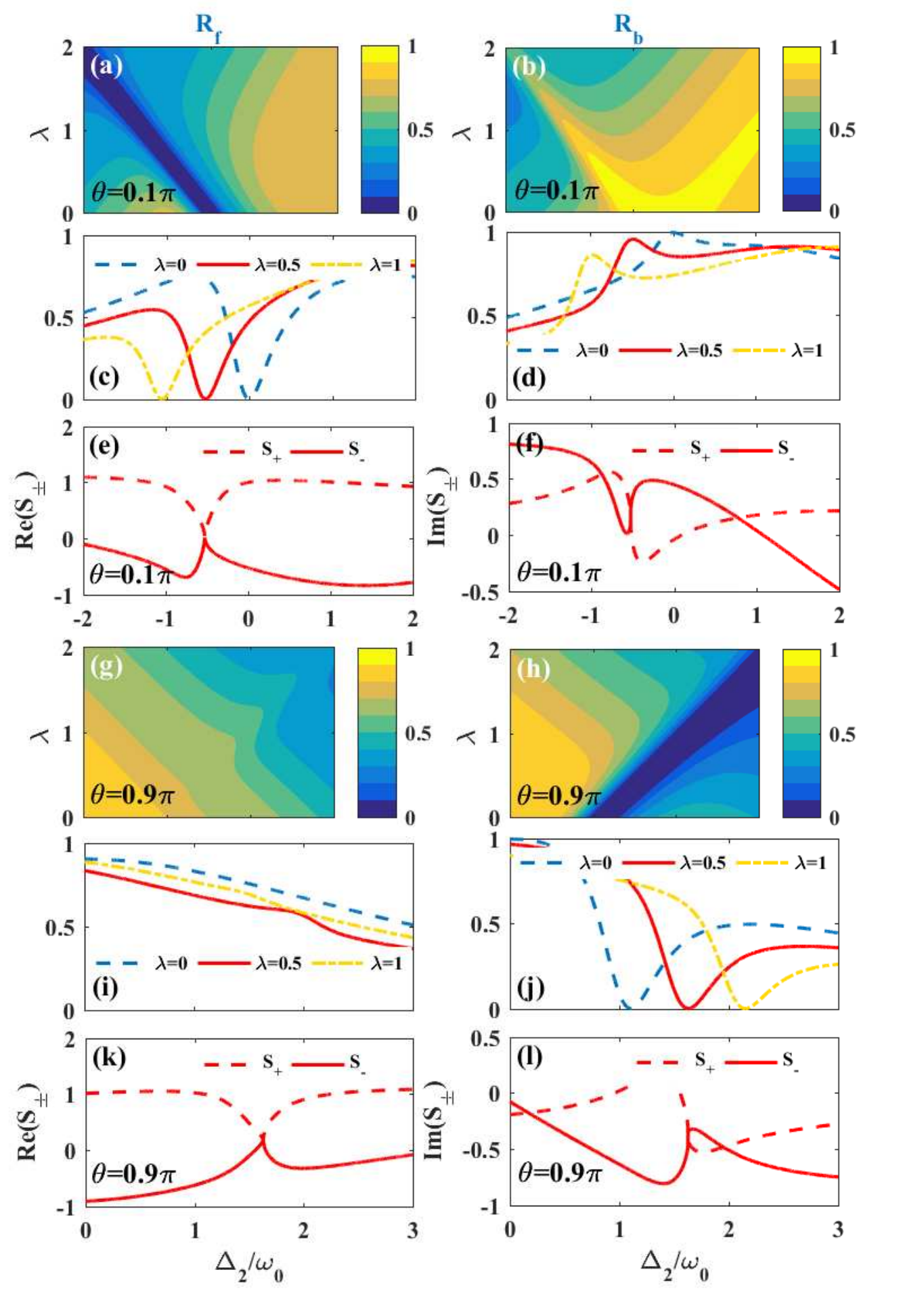}
\caption{(Color online)  (a), (c) and (g), (i) Reflection for forward direction $R_{f}$, (b), (d) and (h), (j) reflection for backward direction $R_{b}$ as a function of coupling strength $\lambda$ and detuning $\Delta_{2}$ for $\theta=0.1\pi$ and $\theta=0.9\pi$, respectively. And (e), (f) and (k), (l) real and imaginary parts of eigenvalues $s_{\pm}$ versus $\Delta_{2}$. Red dashed lines and solid lines represent the eigenvalues $s_{+}$ and $s_{-}$, respectively. Other parameters are $\Gamma= 1.7$, $\gamma_{1} = 0.32$, $\gamma_{2} = 0.01$.} \label{002}
\end{figure}

Secondly,
we discuss the case that
the incident surface plasmon is only resonant with cavity or QD, i.e. $\Delta_{1}=0$ or $\Delta_{2}=0$.
At this point, the two cavity resonators have different transition frequencies.
Similarly, the reflection coefficients are now reduced to
\begin{widetext}
\begin{eqnarray}\label{e8}
t&=&\frac{e^{-i\theta}\{i(-1+e^{2i\theta})\lambda\Gamma+ e^{i\theta} [(\Delta_{1} + i \gamma_{1})(\Delta_{2} + i \gamma_{2})-\lambda^{2}]\}}
{(-1+e^{2i\theta})\Gamma^{2}+i[2e^{i\theta}\lambda\Gamma+\Gamma(\Delta_{1} + i \gamma_{1})+\Gamma(\Delta_{2} + i \gamma_{2})]+[(\Delta_{1} + i \gamma_{1})(\Delta_{2} + i \gamma_{2})-\lambda^{2}]},\\
r_{f}&=&-\frac{(-1+e^{2i\theta})\Gamma^{2}+i[2e^{i\theta}\lambda\Gamma+e^{2i\theta}\Gamma(\Delta_{1} + i \gamma_{1})+\Gamma(\Delta_{2} + i \gamma_{2})]}
{(-1+e^{2i\theta})\Gamma^{2}+i[2e^{i\theta}\lambda\Gamma+\Gamma(\Delta_{1} + i \gamma_{1})+\Gamma(\Delta_{2} + i \gamma_{2})]+[(\Delta_{1} + i \gamma_{1})(\Delta_{2} + i \gamma_{2})-\lambda^{2}]},\\
r_{b}&=&-\frac{(-1+e^{2i\theta})\Gamma^{2}+i[2e^{i\theta}\lambda\Gamma+\Gamma(\Delta_{1} + i \gamma_{1})+e^{2i\theta}\Gamma(\Delta_{2} + i \gamma_{2})]}
{(-1+e^{2i\theta})\Gamma^{2}+i[2e^{i\theta}\lambda\Gamma+\Gamma(\Delta_{1} + i \gamma_{1})+\Gamma(\Delta_{2} + i \gamma_{2})]+[(\Delta_{1} + i \gamma_{1})(\Delta_{2} + i
\gamma_{2})-\lambda^{2}]}.
\end{eqnarray}
\end{widetext}

According to the Eqs.(14)-(15),
we plot the the reflection spectra for forward and backward directions versus coupling strength $\lambda$ and detuning $\Delta_{2}$
with the phase shifts $\theta =0.1\pi$ and $0.9\pi$ in Fig. 6.
For the case of $\theta =0.1\pi$,
the near-zero reflection valleys (blue regions) for forward direction (Fig. 6(a)) have red-shifts and valleys remain as the coupling strength increases (Fig. 6(c)),
and the high reflection peaks (yellow regions) for backward direction (Fig. 6(b))
also undergo red-shifts and the peaks gradually become smaller as the coupling strength increases (Fig. 6(d)).
Namely,
the low reflection region (Fig. 6(a)) correspond to high reflection regions (Fig. 6(b))
in the range of coupling strength from 0 to 0.76 and detuning from -0.74 to 0.1 with high contrast ratio close to 1.
Thence, the unidirectional reflectionlessness can be achieved in the wide range of coupling strength and detuning.
For the case of $\theta =0.9\pi$,
the near-zero reflection regions for backward direction (Fig. 6(g)) with blue-shifts
correspond to high reflection regions for forward direction (Fig. 6(h)) and the contrast ratio close to 0.8.
Under the condition that two phase shifts and coupling strength $\lambda=0.5$,
we draw the real and imaginary parts of eigenvalues $s_{\pm}$ versus $\Delta$  as shown in Figs. 6(e)-(f) and 6(k)-(l), respectively.
The real and imaginary parts of two eigenvalues $s_{+}$ and $s_{-}$ coalesce and cross
at detuning $\Delta=-0.5$ ($\theta = 0.1\pi$) and $\Delta=1.6$ ($\theta =0.9\pi$).
Therefore, unidirectional reflectionless phenomena occur at EPs by advisably adjusting coupling strength.

\begin{figure}[h]
\centering
\includegraphics[angle=0,width=0.5\textwidth]{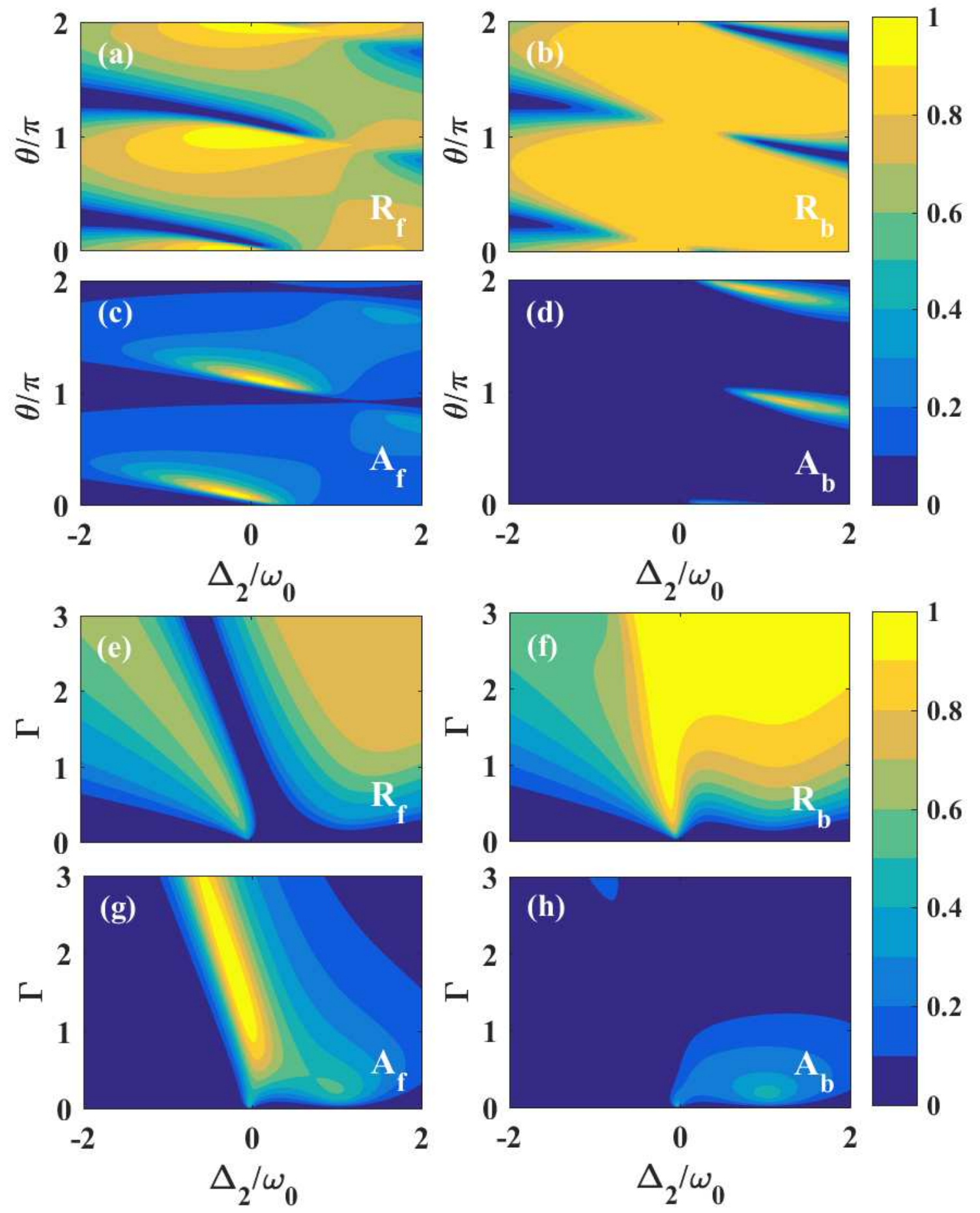}
\caption{(Color online) (a)-(d) Reflections and absorptions for forward and backward directions as a function of phase shift $\theta$ and detuning $\Delta_{2}$. (e)-(h) Reflections and absorptions for forward and backward directions as a function of decay rate $\Gamma$ and detuning $\Delta_{2}$. (a)-(d) $\Gamma= 1.8$, (e)-(h) $\theta=0.1\pi$ and other parameters are the same as in Fig. 4.} \label{03}
\end{figure}

Finally, we investigate the influences of phase shift and decay rate on reflection and absorption in Fig. 7.
From Figs. 7(a) and 7(b),
the low reflection regions (blue regions) for forward direction (Fig. 7(a)) correspond to
high reflection regions (yellow regions) for backward direction (Fig. 7(b))
in a wide range of phase shifts $0.05\pi \sim 0.19\pi$ and $1.05\pi\sim1.23\pi$.
While the low reflection regions (blue regions) for backward direction (Fig. 7(b)) correspond to
high reflection regions (yellow regions) for forward direction (Fig. 7(a))
in phase shifts $0.86\pi\sim0.97\pi$ and $1.86\pi\sim1.96\pi$.
It is obvious that unidirectional reflectionless phenomena change periodically with increasing the phase shift.
Likewise,
the low reflection region (Fig. 7(e)) with a red shift corresponds to the high reflection region (Fig. 7(f))
as the decay rate increases in the range of detuning -0.53$\sim$0.07.
Meanwhile,
the perfect absorption phenomena occur around the phase shifts $\theta=0.1\pi$ and $\theta=1.1\pi$ and the forward absorptivity reaches 1 and backward absorptivity reaches 0.88 around the phase shifts $\theta=0.9\pi$ and $\theta=1.9\pi$ in Figs. 7(c)-7(d), respectively.
Similarly,
perfect absorption for forward direction is also achieved over a wide range of decay rate 0.84$\sim$3 as shown in Figs. 7(g)-7(h).
Thence, bilateral unidirectional reflectionlessness and high absorptivity can be achieved in wide phase shift ranges and decay rate.

\section{Conclusion}\label{Sec:Conclusion}

In conclusion,
we have demonstrated unidirectional reflectionlessness at EPs by modulating the coupling strength between cavity and QD in a waveguide QED system.
When the $\Lambda$-type three-level QD is driven by classical driving field,
dual-band unidirectional reflectionlessness and high absorption can be obtained by properly manipulating the coupling strength, Rabi frequency and phase shift.
In the absence of classical drive,
the first case of the incident surface plasmon resonates with the cavity and QD,
by properly modulating the coupling strength and phase shift,
unidirectional reflectionlessness can be obtained  and the contrast ratio might reach  1.
Meanwhile, the perfect absorption  occurs in wide range of phase shift.
In the second case where the incident surface plasmon is only resonant with cavity or QD,
unidirectional reflectionlessness still remains and the contrast ratio may be larger than 1
in a wide range of coupling strength and phase shift. This is due to the non-hermitian effect of hte sytem.
Moreover, the unidirectional coherent perfect absorption near EPs is obtained
and the absorptivity is about 1.
These results might have the potential applications in nonreciprocal optical invisibility devices,
such as surface plasmon switching, transistor, isolator, diode-like device and so on.

\section*{ACKNOWLEDGMENTS}
This work is supported by National Natural Science Foundation of China (NSFC) under Grants No. $12175033$ and No. $12147206$
and National Key R$\&$D Program of China under Grant No. $2021YFE0193500$.
We thank Yufeng Peng and Dianzhen Cui for helpful discussions and constructive comments in the calculations.

\bibliographystyle{plain}

\begin{thebibliography}{9}

\bibitem{BenderCM1998}
C.M. Bender and S. Boettcher.
Real spectra in non-Hermitian Hamiltonians having PT symmetry.
\href{https://doi.org/10.1103/physrevlett.80.5243.}
{Phys. Rev. Lett. \textbf{80} (1998) 5243.}

\bibitem{JingH2017}
%
H. Jing, S.K. $\rm\ddot{O}$zdemir, H. L$\rm\ddot{u}$ and F. Nori.
High-order exceptional points in optomechanics.
\href{https://doi.org/10.1038/s41598-017-03546-7.}
{Sci. Rep. \textbf{7} (2017) 3386.}
%
\bibitem{PengB2014Science}
%
B. Peng, S.K. $\rm\ddot{O}$zdemir, S. Rotter, H. Yilmaz, M. Liertzer, F. Monifi, C.M. Bender, F. Nori and L. Yang.
Loss-induced suppression and revival of lasingLoss-induced suppression and revival of lasing.
\href{https://doi.org/10.1126/science.1258004.}
{Science \textbf{346} (2014) 328.}
%
\bibitem{PengB2016}
B. Peng, S.K. $\rm\ddot{O}$zdemir, M. Liertzer, W. Chen, J. Kramer, H. Yilmaz, J. Wiersig, S. Rotter and L. Yang.
Chiral modes and directional lasing at exceptional points.
\href{https://doi.org/10.1073/pnas.1603318113.}
{Proc. Natl. Acad. Sci. U. S. A. \textbf{113} (2016) 6845.}
%
\bibitem{DopplerJ2016}
J. Doppler, A.A. Mailybaev, J. B$\rm\ddot{o}$hm, U. Kuhl, A. Girschik, F. Libisch, T.J. Milburn, P. Rabl, N. Moiseyev and S. Rotter.
Dynamically encircling an exceptional point for asymmetric mode switching.
\href{https://doi.org/10.1038/nature18605.}
{Nature \textbf{537} (2016) 76.}
%
\bibitem{PengB2014}
%
B. Peng, S.K. $\rm\ddot{O}$zdemir, F. Lei, F. Monifi, M. Gianfreda, G.L. Long, S. Fan, F. Nori, C.M. Bender and L. Yang.
Parity-time symmetric whispering-gallery microcavities.
\href{https://doi.org/10.1038/nphys2927.}
{Nat. Phys. \textbf{10} (2014) 394.}
%
\bibitem{ChangL2014}
%
L. Chang, X. Jiang, S. Hua, C. Yang, J. Wen, L. Jiang, G. Li, G. Wang and M. Xiao.
Parity-time symmetry and variable optical isolation in active-passive-coupled microresonators.
\href{https://doi.org/10.1038/nphoton.2014.133.}
{Nat. Photonics \textbf{8} (2014) 524.}
%
\bibitem{HeB2018}
%
B. He, L. Yang, X. Jiang and M. Xiao.
Transmission nonreciprocity in a mutually coupled circulating structure.
\href{https://doi.org/10.1103/physrevlett.120.203904.}
{Phys. Rev. Lett. \textbf{120(20)} (2018) 203904.}
%
\bibitem{XiaK2018}
%
K. Xia, F. Nori and M. Xiao.
Cavity-free optical isolators and circulators using a chiral cross-Kerr nonlinearity.
\href{https://doi.org/10.1103/physrevlett.121.203602.}
{Phys. Rev. Lett. \textbf{121(20)} (2018) 203602.}
%
\bibitem{RegensburgerA2012}
%
A. Regensburger, C. Bersch, M.A. Miri, G. Onishchukov, D.N. Christodoulides and U. Peschel.
Parity-time synthetic photonic lattices.
\href{https://doi.org/10.1038/nature11298.}
{Nature \textbf{488} (2012) 167.}
%
\bibitem{JingH2015}
%
H. Jing, S.K. $\rm\ddot{O}$zdemir, Z. Geng, J. Zhang, X.Y. L$\rm\ddot{u}$, B. Peng, L. Yang and F. Nori.
Optomechanically-induced transparency in parity-time-symmetric microresonators.
\href{https://doi.org/10.1038/srep09663.}
{Sci. Rep. \textbf{5} (2015) 9663.}
%
\bibitem{BaumB2015}
%
B. Baum, H. Alaeian and J.A. Dionne.
A parity-time symmetric coherent plasmonic absorber-amplifier.
\href{https://doi.org/10.1063/1.4907871.}
{J. Appl. Phys. \textbf{117} (2015) 063106.}
%
\bibitem{HuangCY2014}
%
C.Y. Huang, R. Zhang, J.L. Han, J. Zheng and J.Q. Xu.
Type-II perfect absorption and amplification modes with controllable bandwidth in combined PT-symmetric and conventional Bragg-grating structures.
\href{https://doi.org/10.1103/physreva.89.023842.}
{Phys. Rev. A \textbf{22} (2014) 023842.}
%
\bibitem{SunY2014}

Y. Sun, W. Tan, H. Li, J. Li and H. Chen.
Experimental demonstration of a coherent perfect absorber with PT phase transition.
\href{https://doi.org/10.1103/physrevlett.112.143903.}
{Phys. Rev. Lett. \textbf{112} (2014) 143903.}
%
\bibitem{LinZ2011}
%
Z. Lin, H. Ramezani, T. Eichelkraut, T. Kottos, H. Cao and D.N. Christodoulides.
Unidirectional invisibility induced by PT-symmetric periodic structures.
\href{https://doi.org/10.1103/physrevlett.106.213901.}
{Phys. Rev. Lett. \textbf{106} (2011) 213901.}
%
\bibitem{AlaeianH2014}
%
H. Alaeian and J.A. Dionne.
Parity-time-symmetric plasmonic metamaterials.
\href{https://doi.org/10.1103/physreva.89.033829.}
{Phys. Rev. A \textbf{89} (2014) 033829.}
%
\bibitem{GearJ2015}
%
J. Gear, F. Liu, S.T. Chu, S. Rotter and J. Li.
Parity-time symmetry from stacking purely dielectric and magnetic slabs.
\href{https://doi.org/10.1103/physreva.91.033825 .}
{Phys. Rev. A \textbf{91} (2015) 033825.}
%
\bibitem{FuY2016}
%
Y. Fu, Y. Xu and H. Chen.
Zero index metamaterials with PT symmetry in a waveguide system.
\href{https://doi.org/10.1364/oe.24.001648 .}
{Opt. Express \textbf{24} (2016) 1648.}
%
\bibitem{FengL2013}
%
L. Feng, X. Zhu, S. Yang, H. Zhu, P. Zhang, X. Yin, Y. Wang and X. Zhang.
Demonstration of a large-scale optical exceptional point structure.
\href{https://doi.org/10.1364/oe.22.001760 .}
{Opt. Express \textbf{22} (2013) 1760.}
%
\bibitem{YangE2016}

E. Yang, Y. Lu, Y. Wang, Y. Dai and P. Wang.
Unidirectional reflectionless phenomenon in periodic ternary layered material.
\href{https://doi.org/10.1364/oe.24.014311 .}
{Opt. Express \textbf{24} (2016) 14311.}
%
\bibitem{ShenY2014}
%
Y. Shen, X. H. Deng and L. Chen.
Unidirectional invisibility in a two-layer non-PT-symmetric slab.
\href{https://doi.org/10.1364/oe.22.019440 .}
{Opt. Express \textbf{22} (2014) 19440.}
%
\bibitem{Gu2017OE}
%
X. Gu, R. Bai, C. Zhang, X.R. Jin, Y.Q. Zhang, S. Zhang and Y.P. Lee.
Unidirectional reflectionless propagation in a non-ideal parity-time metasurface based on far field coupling.
\href{ https://doi.org/10.1364/oe.25.011778 .}
{Opt. Express \textbf{25} (2017) 11778.}
%
\bibitem{Bai2017SR}
%
R. Bai, C. Zhang, X. Gu, X.R. Jin, Y.Q. Zhang and Y.P. Lee.
Switching the unidirectional refectionlessness by polarization in non-ideal PT metamaterial based on the phase coupling.
\href{https://doi.org/10.1038/s41598-017-11376-w.}
{Sci. Rep. \textbf{7} (2017) 10742.}
%
\bibitem{Bai2017APE}

R. Bai, C. Zhang, X. Gu, X.R. Jin, Y.Q. Zhang and Y.P. Lee.
Unidirectional reflectionlessness and perfect nonreciprocal absorption in stacked asymmetric metamaterial based on near-field coupling.
\href{ https://doi.org/10.7567/apex.10.112001 .}
{Appl. Phys. Express \textbf{10} (2017) 112001.}
%
\bibitem{Yin2018OC}
%
H. Yin, R. Bai, X. Gu, C. Zhang, G.R. Gu, Y.Q. Zhang, X.R. Jin and Y.P. Lee.
Unidirectional reflectionless propagation in non-Hermitian metamaterial based on phase coupling between two resonators.
\href{https://doi.org/10.1016/j.optcom.2018.01.019 .}
{Opt. Commun. \textbf{414} (2018) 172.}
%
\bibitem{Liu2018JNOPM}
%
S.J. Liu, C. Zhang, R. Bai, X. Gu, H.D. Yin, Y.H. Liu, Y.Q. Zhang, X.R. Jin and Y.P. Lee.
Unidirectional reectionlessness and nonreciprocal perfect absorption in optical waveguide system based on phase coupling between two photonic crystal cavities.
\href{https://doi.org/10.1142/s0218863518500042 .}
{J. Nonlinear Opt. Phys. Mater. \textbf{27} (2018) 1850004.}
%
\bibitem{HuangY2016}
%
Y. Huang, C. Min and G. Veronis.
Broadband near total light absorption in non-PT-symmetric waveguide-cavity systems.
\href{https://doi.org/10.1364/oe.24.022219 .}
{Opt. Express \textbf{24} (2016) 22219.}

%
\bibitem{ZhangC2017}
%
C. Zhang, R. Bai, X. Gu, Y. Jin, Y.Q. Zhang, X.R. Jin, S. Zhang and Y.P. Lee.
Unidirectional reflectionless phenomenon in ultracompact non-Hermitian plasmonic waveguide system based on phase coupling.
\href{https://doi.org/10.1088/2040-8986/aa9532 .}
{J. Opt. \textbf{19} (2017) 125005.}
%
\bibitem{ZhangC2017IEEE}

C. Zhang, R. Bai, X. Gu, X.R. Jin, Y.Q. Zhang and Y.P. Lee.
Unidirectional reflectionless propagation in plasmonic waveguide system based on phase coupling between two stub resonators.
\href{https://doi.org/10.1109/jphot.2017.2761899 .}
{IEEE Photonics J. \textbf{9} (2017) 1.}
%
\bibitem{ZhangC2017OE}

C. Zhang, R. Bai, X. Gu, X.R. Jin, Y.Q. Zhang and Y.P. Lee.
Dual-band unidirectional reflectionless phenomena in an ultracompact non-Hermitian plasmonic waveguide system based on near-field coupling.
\href{https://doi.org/10.1364/oe.25.024281 .}
{Opt. Express \textbf{25} (2017) 24281.}
%
\bibitem{ShenJT2007}
%
J.-T. Shen and S. Fan.
Strongly correlated two-photon transport in a one-dimensional waveguide coupled to a two-level system.
\href{https://doi.org/10.1103/physrevlett.98.153003 .}
{Phys. Rev. Lett. \textbf{98} (2007) 153003.}
%
\bibitem{Fan Shanhui2010}
%
S. Fan, S.E. Kocabas and J.-T. Shen.
Input-output formalism for few-photon transport in one-dimensional nanophotonic waveguides coupled to a qubit.
\href{https://doi.org/10.1103/physreva.82.063821 .}
{Phys. Rev. A \textbf{82} (2010) 063821.}
%
\bibitem{Rephaeli Eden2011}
%
E. Rephaeli, S.E. Kocabas and S. Fan.
Few-photon transport in a waveguide coupled to a pair of colocated two-level atoms.
\href{https://doi.org/10.1103/physreva.84.063832 .}
{Phys. Rev. A \textbf{84} (2011) 063832.}
%
\bibitem{Longo Paolo2011}
%
P. Longo, P. Schmittrckert and K. Busch.
Few-photon transport in low-dimensional systems.
\href{https://doi.org/10.1103/physreva.83.063828 .}
{Phys. Rev. A \textbf{83} (2011) 063828.}
%
\bibitem{Zheng Huaixiu2012}
%
H. Zheng, D.J. Gauthier and H.U. Baranger.
Strongly correllated photons generated by coupling a three- or four-level system to a waveguide.
\href{https://doi.org/10.1103/physreva.85.043832 .}
{Phys. Rev. A \textbf{85} (2012) 043832.}
%
\bibitem{Roy Dibyendu2013}
%
D. Roy.
Two-photon scattering of a tightly focused weak light beam from a small atomic ensemble:
An optical probe to detect atomic level structures.
\href{https://doi.org/10.1103/physreva.87.063819 .}
{Phys. Rev. A \textbf{87} (2013) 063819.}
%
\bibitem{Englund Dirk2010}
%
D. Englund, A. Majumdar, A. Faraon, M. Toishi, N. Stoltz, P. Petroff and J. Vu\v{c}kovi\'{c}.
Resonant excitation of a quantum dot strongly coupled to a photonic crystal nanocavity.
\href{https://doi.org/10.1103/physrevlett.104.073904 .}
{Phys. Rev. Lett. \textbf{104} (2010) 073904.}
%

\bibitem{MTChen2012OL}
%
M.T. Cheng and Y.Y. Song.
Fano resonance analysis in a pair of semiconductor quantum dots coupling to a metal nanowire.
\href{https://doi.org/10.1364/ol.37.000978 .}
{Opt. Lett. \textbf{37} (2012) 978.}
%
\bibitem{Huang Jin-Feng2013}
%
J.-F. Huang, T. Shi, C.P. Sun and F. Nori.
Controlling single-photon transport in waveguides with finite cross section.
\href{https://doi.org/10.1103/physreva.88.013836 .}
{Phys. Rev. A \textbf{88} (2013) 013836.}
%
\bibitem{Liao Jie-Qiao2013}
%
J.-Q. Liao and C.K. Law.
Correlated two-photon scattering in cavity optomechanics.
\href{https://doi.org/10.1103/physreva.87.043809 .}
{Phys. Rev. A \textbf{87} (2013) 043809.}
%
\bibitem{Shi T2011}
%
T. Shi, S. Fan and C.P. Sun.
Two-photon transport in a waveguide coupled to a cavity in a two-level system.
\href{https://doi.org/10.1103/physreva.84.063803 .}
{Phys. Rev. A \textbf{84} (2011) 063803.}
%
\bibitem{JTshen2009PRA}
%
J.T. Shen and S. Fan.
Theory of single-photon transport in a single-mode waveguide. II. Coupling to a whispering-gallery resonator containing a two-level atom.
\href{https://doi.org/10.1103/physreva.79.023838 .}
{Phys. Rev. A \textbf{79} (2009) 023838.}
%
\bibitem{Yan C2011}
%
C.-H. Yan, L.-F. Wei, W.-Z. Jia and J.-T. Shen.
Controlling resonant photonic transport a long optical waveguides by two-level atoms.
\href{https://doi.org/10.1103/physreva.84.045801 .}
{Phys. Rev. A \textbf{84} (2011) 045801.}
%
\bibitem{Shomroni2014}
%
I. Shomroni, S. Rosenblum, Y. Lovsky, O. Bechler, G. Guendelman and B. Dayan.
All-optical routing of single photons by a one-atom switch controlled by a single photon.
\href{https://doi.org/10.1126/science.1254699 .}
{Science \textbf{345} (2014) 903.}
%
\bibitem{Witthaut2010}
%
D. Witthaut and A.S. S{\o}rensen.
Photon scattering by a three-level emitter in a one-dimensional waveguide.
\href{https://doi.org/10.1088/1367-2630/12/4/043052 .}
{New J. Phys. \textbf{12} (2010) 043052.}
%
\bibitem{Gonzalez-Ballestero Carlos2016}
%
C. Gonzalez-Ballestero, E. Moreno, F.J. Garcia-Vidal and A. Gonzalez-Tudela.
Nonreciprocal few-photon routing schemes based on chiral waveguide-emitter couplings.
\href{https://doi.org/10.1103/physreva.94.063817 .}
{Phys. Rev. A \textbf{94} (2016) 063817.}
%
\bibitem{Yan Cong-Hua2018}
%
C.-H. Yan, Y. Li, H. Yuan and L.F. Wei.
Targeted photonic routers with chiral photon-atom interactions.
\href{https://doi.org/10.1103/physreva.97.023821 .}
{Phys. Rev. A \textbf{97} (2018) 023821.}
%
\bibitem{RoyD2011}
%
D. Roy.
Two-photon scattering by a driven three-level emitter in a one-dimensional waveguide and electromangetically induced transparency.
\href{https://doi.org/10.1103/physrevlett.106.053601 .}
{Phys. Rev. Lett. \textbf{106} (2011) 053601.}
%
\bibitem{Xu2017}
%
X.-W. Xu, A.-X. Chen, Y. Li and Y. Liu.
Single-photon nonreciprocal transport in one-dimensional coupled-resonator waveguides.
\href{https://doi.org/10.1103/physreva.95.063808 .}
{Phys. Rev. A \textbf{95} (2017) 063808.}
%
\bibitem{Shangcheng2019}
%
C. Shang, H.Z. Shen and X.X. Yi.
Nonreciprocity in a strongly coupled three-mode optomechanical circulatory system.
\href{https://doi.org/10.1364/oe.27.025882 .}
{Opt. Express \textbf{27} (2019) 25882.}
%
\bibitem{Nie2021}
%
W. Nie, T. Shi, F. Nori and Y. Liu.
Topology-enhanced nonreciprocal scattering and photon absorption in a waveguide.
\href{https://doi.org/10.1103/physrevapplied.15.044041 .}
{Phys. Rev. Appl. \textbf{15} (2021) 044041.}
%
\bibitem{Huang2015}
%
Y. Huang, G. Veronis and C. Min.
Unidirectional reflectionless propagation in plasmonic waveguide-cavity systems at exceptional points.
\href{https://doi.org/10.1364/oe.23.029882 .}
{Opt. Express \textbf{23} (2015) 029882.}
%
\bibitem{Wu2018OE}
%
N. Wu, C. Zhang, X.R. Jin, Y.Q. Zhang and Y.P. Lee.
Unidirectional reflectionless phenomena in non-Hermitian quantum system of quantum dots coupled to a plasmonic waveguide.
\href{https://doi.org/10.1364/oe.26.003839 .}
{Opt. Express \textbf{26} (2018) 3839.}
%
\bibitem{Qiu2019QIP1}

D.-X. Qiu, R. Bai, C. Zhang, L.F. Xin, X.Y. Zou, Y.Q. Zhang, X.R. Jin, C. An and S. Zhang.
Unidirectional reflectionlessness in a non-Hermitian quantum system of surface plasmon coupled to two plasmonic cavities.
\href{https://doi.org/10.1007/s11128-018-2139-8 .}
{Quant. Inf. Process. \textbf{18} (2019) 28.}
%
\bibitem{Qiu2019QIP2}
%
D.-X. Qiu, X.Y. Zou, Y. Liu, H. Yang, L. Yu, Y.Q. Zhang, X.R. Jin, C. An and S. Zhang.
Dual-band unidirectional reflectionlessness in non-Hermitian quantum system consisting of a gain and a loss plasmonic cavities.
\href{https://doi.org/10.1007/s11128-019-2380-9 .}
{Quant. Inf. Process. \textbf{18} (2019) 269.}
%
\bibitem{Yang2020QST}
%
H. Yang, D.-X. Qiu, X.Y. Zou, C. An, Y.Q. Zhang and X.R. Jin.
Active controlled dual-band unidirectional reflectionlessness by classical driving field in non-Hermitian quantum system.
\href{https://doi.org/10.1088/2058-9565/abb4a7 .}
{Quantum Sci. Technol. \textbf{5} (2020) 045018.}
%
\bibitem{PRB842011}
%
G.Y. Chen, N. Lambert, C.H. Chou, Y.N. Chen and F. Nori.
Surface plasmons in a metal nanowire coupled to colloidal quantum dots: scattering properties and quantum entanglement.
\href{https://doi.org/10.1103/physrevb.84.045310 .}
{Phys. Rev. B \textbf{84} (2011) 045310.}
%
\bibitem{104}
%
L. Feng, Y.L. Xu, W.S. Fegadolli, M.H. Lu, J.E.B. Oliveria, V.R. Almeida, Y.F. Chen and A. Scherer.
Experimental demonstration of a unidirectional reflectionless parity-time metamaterial at optical frequencies.
\href{https://doi.org/10.1038/nmat3495 .}
{Nat. Mater. \textbf{12} (2013) 108.}
%
\bibitem{JTShen2005OL}
%
J.T. Shen and S. Fan.
Coherent photon transport from spontaneous emission in one-dimensional waveguide.
\href{https://doi.org/10.1364/ol.30.002001 .}
{Opt. Lett. \textbf{30} (2005) 2001.}
%
\bibitem{HZheng2013PRL}
%
H. Zheng and H.U. Baranger.
Persistent quantum beats and long-distance entanglement from waveguide-mediated interactions.
\href{https://doi.org/10.1103/physrevlett.110.113601 .}
{Phys. Rev. Lett. \textbf{110} (2013) 113601.}
%
\bibitem{FFratini2014PRL}
%
F. Fratini, E. Mascarenhas, L. Safari, J-Ph. Poizat, D. Valente, A. Auff\`{e}ves, D. Gerace and M.F. Santos.
Fabry-Perot interferometer with quantum mirrors: nonlinear light transport and rectification.
\href{https://doi.org/10.1103/physrevlett.113.243601 .}
{Phys. Rev. Lett. \textbf{113} (2014) 243601.}

\end{thebibliography}

\end{document}